# Silicates on Iapetus from Cassini's Composite Infrared Spectrometer

Cindy L. Young[1,2], James J. Wray[1], Roger N. Clark[3], John R. Spencer[4], Donald E. Jennings[5]
Kevin P. Hand[6], Michael J. Poston[7], Robert W. Carlson[6]

[1]School of Earth and Atmospheric Sciences, Georgia Institute of Technology, Atlanta, GA

[2]Emory University, Atlanta, GA, USA

[3]Planetary Science Institute, Tucson, AZ, USA

[4]Southwest Research Institute, Boulder, CO, USA

[5]NASA Goddard Space Flight Center, Greenbelt, MD, USA

[6]Jet Propulsion Laboratory, Pasadena, CA, USA

[7]Caltech, Pasadena, CA, USA

**Abstract**

We present the first spectral features obtained from Cassini's Composite Infrared Spectrometer (CIRS) for any icy moon. The spectral region covered by CIRS focal planes (FP) 3 and 4 is rich in emissivity features, but previous studies at these wavelengths have been limited by low signal to noise ratios (S/Rs) for individual spectra. Our approach is to average CIRS FP3 spectra to increase the S/R and use emissivity spectra to constrain the composition of the dark material on Iapetus. We find an emissivity feature at ~855 cm$^{-1}$ and a possible doublet at 660 and 690 cm$^{-1}$ that do not correspond to any known instrument artifacts. We attribute the 855 cm$^{-1}$ feature to fine-grained silicates, similar to those found in dust on Mars and in meteorites, which are nearly featureless at shorter wavelengths. Silicates on the dark terrains of Saturn's icy moons have been suspected for decades, but there have been no definitive detections until now. Serpentines reported in the literature at ambient temperature and pressure have features near 855 and 660 cm$^{-1}$. However, peaks can shift depending on temperature and pressure, so measurements at Iapetus-like conditions are necessary for more positive feature identifications. As a first investigation, we measured muscovite at 125K in a vacuum and found that this spectrum does match the emissivity feature near 855 cm$^{-1}$ and the location of the doublet. Further measurements are needed to robustly identify a specific silicate, which would provide clues regarding the origin and implications of the dark material.

## 1. Introduction

Spectroscopy of icy satellite surfaces can aid us in understanding sources and sinks of material in the outer solar system. These worlds are composed of much more than just $H_2O$ ice, with tantalizing evidence from Cassini's Visual and Infrared Mapping Spectrometer (VIMS) for $CO_2$, possible $NH_3$ and hydrous minerals, organics, and metallic and oxidized Fe on Saturn's icy moons (e.g., Clark et al. 2005, 2012). These non-$H_2O$ constituents appear to be concentrated in the darker terrains of Saturn's outer icy moons: Phoebe, Iapetus, Hyperion, Rhea, and Dione (Clark et al. 2008; Stephan et al. 2012). The dark material may have resulted from endogenic processes (Smith et al. 1982) or exogenic sources, originating from either within (e.g., Soter 1974; Buratti et al. 2002) or outside the Saturnian system (e.g., Clark et al. 2008; Stephan et al. 2010), the latter of which is currently the most widely accepted explanation. For decades, it was speculated that the dark material contained silicates (Lebofsky et al. 1982; Jarvis et al. 2000). A phyllosilicate (Clark et al. 2005) and/or space weathered silicate (Clark et al. 2014) interpretation is consistent with VIMS spectra, yet there has been no definitive detection of silicates until now.

The spectral complexity of the Saturnian satellite system as seen in reflected sunlight suggests that additional complexity may be present at mid-infrared wavelengths detectable via thermal emission, from which unique compositional information can be gleaned (Flasar et al. 2004). However, to date, Cassini Composite Infrared Spectrometer (CIRS) surface compositional studies have received less attention than measurements of atmospheres, surface temperatures, and thermophysical properties across the Saturnian system. Carvano et al. (2007) calculated global average emissivity spectra for five moons (Phoebe, Iapetus, Enceladus, Tethys, and Hyperion), focusing exclusively on the far-infrared portion of the spectrum collected by CIRS focal plane 1 (FP1), which spans 10–600 $cm^{-1}$ (16.7–1000 μm). Relative to their uncertainties

(typically a few percent in emissivity), no spectral features were observed in these global averages. Howett et al. (2014) similarly reported no emissivity features in the FP1 range on the cold polar regions of Rhea and Dione.

The CIRS icy satellite data set has grown substantially since Carvano et al. (2007) completed their work, with close flybys of several moons providing opportunities now to search for spatially heterogeneous surface spectral features, and/or to average more spectra to further reduce noise-related uncertainties. In addition, the mid-IR portion of the CIRS spectrum captured by focal planes 3 and 4 (FP3 and FP4), with spectral ranges of 600–1100 cm$^{-1}$ (9.1–16.7 µm) and 1100–1400 cm$^{-1}$ (7.2–9.1 µm), respectively, remains largely unexplored for icy satellites. The spectral region covered by FP3 and FP4 is rich in emissivity features due to both simple and complex molecules (e.g., Hand et al. 2009), and the strongest fundamental vibrational features of silicates occur in the CIRS wavelength range (Salisbury et al. 1991; Christensen et al. 2004). However, the study of emissivity variations in this region is often challenged by low signal-to-noise ratios (S/Rs) for individual spectra. Here, we focus on FP3 because the wavelength of peak energy emission is longer at the colder temperatures observed on icy moons. Our first attempt at finding spectral features is limited to the dark material on Iapetus, which is comparatively warmer than other icy moon surfaces.

In VIMS spectra, the dark material on Iapetus has shown evidence for nanophase metallic iron, and also exhibits ultraviolet and visible-wavelength absorptions consistent with fine-grained hematite, $Fe_2O_3$ (Clark et al. 2012). $NH_3$ was tentatively identified in VIMS spectra of Iapetus, Phoebe, and Dione (Clark et al. 2008). $CO_2$ was also found in the dark terrains on Iapetus and was interpreted as a product of irradiating $H_2O$ + carbonaceous material such as hydrocarbons (Buratti et al. 2005; Palmer & Brown 2011). Similar ice "contaminants" have been found on Hyperion (Tosi et al. 2010; Dalton et al. 2012), consistent with a common source of surface material for Saturn's three mid-sized moons orbiting beyond Titan. The later discovery of similar dark material on Dione and its spatial distribution there supported a hypothesis that this material arrived at all of these satellites from a source outside the Saturnian system (Clark et al. 2008; Stephan et al. 2010), while the discovery of the Phoebe ring in 2009 led to conclusions that dark particles were transported to Iapetus and Hyperion from Phoebe (Tosi et al. 2010).

The goal of this work is to average CIRS FP3 spectra to increase the S/R and use emissivity spectra to constrain the composition of the dark material on Iapetus. We present the first CIRS spectral feature(s) from an icy moon surface and make efforts to identify the features based on laboratory analysis of terrestrial analog samples and similar mid-IR features observed on other planetary bodies (e.g., planets and meteorites) from the literature. Our results provide clues regarding the origin and transport of exogenous dark material within the Saturnian system.

## 2. Observations and laboratory procedures

CIRS FP3 calibrated spectra from the Planetary Data System (PDS) were obtained using the Vanilla software package (http://software.mars.asu.edu/vanilla/). The Outer Planets Unified Search (OPUS; (http://pds-rings-tools.seti.org/opus/) was used to select observations during which the dark material on Iapetus was illuminated by afternoon sunlight in order to assure the highest surface temperatures possible, and therefore the highest S/R. OPUS preview images (Fig. 1) were used to determine that the observations include good coverage over the dark terrain. Spectra for which the target was not in the field of view, all parts of the detector did not intersect the target, and/or the shutter was not open were removed from the data set. We selected regions covering the thickest parts of the dark deposits, avoiding edges of the dark material and containing very few optically bright features.

Boxed domains for three observations were selected (Fig. 2). There were 46 spectra collected in the orange box, five spectra in the purple box, and one spectrum in the red box. Spectra inside the boxed regions were averaged, and then the wavenumber form of the Plank blackbody equation was used to determine surface temperatures for the average. We expect that based on the close geographic proximity of all spectra included in the averages for the same observation and the continuity of the dark material deposits in these regions that temperatures among the averaged spectra were likely very similar. Although averaging spectra to improve S/R was key in this study, its overuse may be detrimental. For example, for Iapetus, Carvano et al. (2007) studied only a global average including both bright and dark terrains, which have very different reflectance spectral properties. Therefore, care was taken not to over-average spectra so that spectral features become indistinguishable. Emissivities were calculated by dividing the CIRS average radiances by the blackbody radiance at each wavenumber. We identified possible spectral features by examining both radiance and emissivity curves for "bumps" that deviated from—and then returned to—the continuum.

Comprehensive infrared spectral libraries of minerals are widely available, but most of these data were acquired at "inner solar system" (usually Earth-like) conditions. Spectral features for materials, such as silicates, can shift substantially in position and relative strength as a function of changing temperature (e.g., Dalton et al. 2005; Wray et al. 2014) or ambient pressure (Logan et al. 1973; Wray et al. 2014). Grain size also dramatically influences mid-IR spectra (Wray et al. 2014), with sub-micron grains as suggested by Iapetus' low thermal inertia (Howett et al. 2010; Rivera-Valentin et al. 2011), appearing remarkably different due to the enhancement of transparency features (~ 11 – 14 µm or 710 – 910 cm$^{-1}$) and the disappearance of reststrahlen bands (~ 9 – 12 µm or 830 – 1100 cm$^{-1}$) (Logan et al. 1973). Prior spectral measurements of silicates have largely focused on coarser-grained samples in an effort to minimize the transparency features of prime interest to us.

The Icy Worlds Simulation Lab at JPL contains two ultra-high vacuum chambers each equipped with a cryostat (20 – 300 K), electron gun, and MIDAC Fourier Transform IR spectrometer. An additional spectrometer covers the visible and shortwave IR, allowing for continuous spectral coverage from 0.28 to ~20 µm. A Diffuse aluminum target was measured and served as a background for the sample of interest. We initially collected spectra under ambient air in a purged N$_2$ glovebox in order to confirm transparency features (indicating sufficiently fine grain sizes) and identify the best candidate minerals for more time-consuming cold vacuum measurements. Out of several silicate samples (i.e., FeSiO "smoke," albite, diopside, muscovite, quartz, and labradorite), muscovite provided the closest match under ambient air conditions to CIRS spectral features and was therefore selected as the first mineral on which to take the longer duration cold vacuum measurements. Muscovite was ground and sieved to < 0.4 mm, then vibrated in a wet micronizing mill for 20 minutes to generate very fine-grained particles, on the order of a few microns (O'Connor & Chang, 1986), in a slurry. Water was evaporated off, yet remnant thin films may have contributed to the observed clumping of the µm-scale grains into ~0.2 mm aggregates. Micronized muscovite was measured at 125 K in a vacuum (2.7x10$^{-8}$ torr) with a 32º phase angle. The sample, source, and detector were configured in a directional, bi-conical geometry. Emissivity was calculated as one minus the measured spectral reflectance.

**3. Observational results**

Fig. 3a illustrates the benefits of spectral averaging. Noise features are visible at all but the lowest wavenumbers in five individual detector spectra from ~7°S, 220°W, but are mitigated by averaging just these five detectors (Fig. 3b). However, the strongest "features" in the five-detector average (near 770 and 960 cm$^{-1}$) are still artifacts, as illustrated by their presence in CIRS spectra of empty deep space (Fig. 3c). Such artifacts vary in position from one spectrum to another, and therefore can be eliminated by averaging all of the spectra within a larger region. Therefore, all spectra obtained within each of the boxed regions (Fig. 2) were averaged. The resulting radiance spectra for each region have greatly reduced noise and artifacts (Fig. 3c). A consistent dip near ~855 cm$^{-1}$ seen in the red and purple regions does not correspond to any artifact.

Averaging 46 spectra from the orange boxed domain from Fig. 2 yields a featureless radiance curve (Fig. 3c) and an emissivity spectrum appearing relatively noise-free below ~930 cm$^{-1}$ (Fig. 4). An emissivity feature of interest at ~855 cm$^{-1}$ (11.7 μm) dips ~4% below the baseline and returns back to baseline prior to the onset of noise at the highest wavenumbers. By contrast, pure H$_2$O ice at these temperatures has a broad emissivity peak from ~750 to ~950 cm$^{-1}$ (Allodi et al. 2014), so our feature strength may be >5% if mixed with substantial H$_2$O ice. This feature does not correspond to any known instrument artifact, and the standard deviation of the averaged spectra shows no unusual behavior at this wavelength. To the best of our knowledge, this is the first spectral feature reported from CIRS for an icy moon. There is also a pair of possible features near 660 and 690 cm$^{-1}$ (~15 μm), which appear to be above the noise levels in this part of the spectrum, albeit weaker than the feature reported at 855 cm$^{-1}$.

**4. Spectral feature identification and discussion**

In order to identify the spectral features, analog materials must be analyzed at both CIRS and VIMS spectral ranges. An initial comparison of the CIRS features to materials measured by the United States Geological Survey at Earth ambient temperature and pressure was useful in narrowing down the candidates to measure at Iapetus-like conditions. In doing this, we see that elemental sulfur does have a reflectance peak (emissivity minimum) near 855 cm$^{-1}$ and even has another weaker feature at ~660 cm$^{-1}$ as in our CIRS spectrum (Clark et al. 2007), but also has a distinct yellow color caused by an abrupt drop to near-zero reflectance below ~0.5 μm, which VIMS has not observed in Iapetus' dark terrains (e.g., Clark et al. 2012). Carbonates, such as siderite ($FeCO_3$) or trona ($Na_3[CO_3][HCO_3]\cdot 2H_2O$), have fundamental $CO_3^{2-}$ bend vibrations at ~870 cm$^{-1}$ (Clark 1999), but also have shortwave IR absorptions that VIMS has not observed on Iapetus.

By contrast, there are vast dust-covered regions of Mars across which the only feature at CIRS FP3 wavelengths is an emissivity minimum at ~840 cm$^{-1}$ (Fig. 4); at VIMS wavelengths, only nanophase $Fe^{3+}$ and minor $H_2O$ (both also present on Iapetus) are detected in these Martian regions. The ~840 cm$^{-1}$ feature on Mars is attributed to fine-grained silicate minerals such as feldspars (Christensen et al. 2004). Very fine grains are required to produce this "transparency feature" along with a lack of silicate reststrahlen bands. If silicates are present on Iapetus, then the extremely low thermal inertia values measured there—between 6 and 21 J m$^{-2}$ K$^{-1}$ s$^{-\frac{1}{2}}$ on the leading side (Howett et al. 2010; Rivera-Valentin et al. 2011), with the uppermost surface <10 J m$^{-2}$ K$^{-1}$ s$^{-\frac{1}{2}}$ (Neugebauer et al. 2005)—are consistent with sub-micron-size grains (Presley & Christensen 1997). However, if the abundance of silicates is low, then they may contribute only negligibly to the total near-surface thermal inertia. Coarser-grained silicates would have an additional, deeper (reststrahlen) emissivity minimum at ~950–1100 cm$^{-1}$ (e.g., Salisbury et al.

1991), which would be consistent with our CIRS spectrum but cannot be confirmed due to our lower S/R in this spectral region.

A further comparison to published data suggests that the best match to the feature observed at 855 cm$^{-1}$ at ambient air temperature and pressure is altered magnesium-rich phyllosilicates, in particular, serpentines such as antigorite and lizardite (Figure 4; Bishop et al. 2008a, 2008b). In addition, the large dip in the rather noisy part of the CIRS spectra is consistent with the reststrahlen bands of the serpentines near ~1045 cm$^{-1}$. The antigorite and lizardite spectra in Fig. 4 also match the ~660 cm$^{-1}$ part of the doublet observed in the CIRS emissivity spectrum (Fig. 4); although a doublet is not observed here in antigorite and lizardite measured at room temperature, lower temperatures often resolve fine structure within bands that appear broad and merged at room temperature (e.g., Dalton et al. 2005). Another possible explanation for the dips at ~660 cm$^{-1}$ is $CO_2$ ice, which should have a singlet or doublet there due to the O-C-O bending mode (White et al. 2009; Allodi et al. 2014). Iapetus has the strongest $CO_2$ features among Saturn's moons at VIMS wavelengths (Pinilla-Alonso et al., 2011), concentrated in the dark terrains that we analyzed with CIRS (Buratti et al. 2005; Cruikshank et al. 2010).

The central wavelength of the silicate transparency feature shifts with silicate mineralogy (Salisbury & Walter 1989), so we must constrain the nature of Saturnian system silicates by comparing CIRS spectra to those of various silicates measured at low temperatures and vacuum conditions. Muscovite was a good match to the CIRS data with a feature at ~850 cm$^{-1}$ (Fig. 4); it is a phyllosilicate similar to those tentatively found on Europa and attributed to impact delivery (Shirley et al. 2013). VIMS has not detected clear features of muscovite on Iapetus in its spectral range (Clark et al. 2012), but we note that meteorites have been observed to be similarly featureless in the shortwave IR while nonetheless retaining silicate reflectance peaks in the mid-

IR (e.g., McAdam et al. 2015). Emissivity minima in the noisy parts of the muscovite spectrum are also consistent with potential CIRS features at ~ 660 and 690 cm$^{-1}$. In addition, recent laboratory spectral modeling shows that space-weathered silicates provide a good match to the VIMS visible and shortwave IR data from Iapetus (Clark et al. 2014). Regardless, we emphasize that it would be premature to attribute any icy moon emissivity features to a specific silicate at this point.

Silica ($SiO_2$) nanoparticles have recently been reported in the Saturn system, specifically in the E ring (Sekine et al. 2013). Cassini's Cosmic Dust Analyzer (CDA) has previously detected silicates elsewhere in the Saturn system: Kempf et al. (2005) reported "streams of tiny dust particles escaping from the Saturnian system" that "consist predominantly of oxygen, silicon, and iron." Hsu et al. (2015) argued that the Enceladus plume is the source of these silicates and that they are a product of cryovolcanic/hydrothermal activity from within the moon's ocean. Alternatively, since dark material on Iapetus and other moons is thought to be transported from the outer captured moon Phoebe, or from another source outside the Saturn system (e.g., Cruikshank et al. 1983), putative silicates on Iapetus could be meteoritic in origin. In particular, serpentines, such as antigorite, have been detected on meteorites (McAdam et al. 2015) and are a product of hydrothermal activity on the Earth (Martin et al. 2008). However, an endogenic source or some combination of internal and external processes for the formation of the dark material cannot be completely ruled out (Smith et al. 1982).

Silicates were previously undetected on Saturnian moons, but have been suggested as possible components of the dark material (e.g., Lebofsky et al. 1982; Clark et al. 2005). The particles would darken the ice surfaces and alter the surface radiation balance, similar to deposited ash from volcanoes on Earth (Flanner et al. 2014; Young et al. 2014). This absorbing dark surface combined with Iapetus' slow rotation can cause the ice to heat, sublime, and migrate (Spencer 1987). Additionally, Earth's volcanoes and hydrothermal systems have been credited with providing the early planet with the materials needed for life (Johnson et al. 2008; Martin et al. 2008). Although the chemical composition of the Enceladan plume has yet to be fully characterized, a volcanic source for the dark material would hint that bio-essential nutrients may have been delivered to the surfaces of the darkened Saturnian bodies and potentially also to Titan, which orbits in their midst. A more specific mineral identity for Iapetus' putative silicates would help to establish their origin and implications.

This work was supported by NASA Outer Planets Research Program grant NNX14AO34G and by Georgia Tech's Center for Space Technology and Research (C-STAR). The authors thank Janice Bishop for providing serpentine lab spectra and Richard Cartwright, Tim Glotch, and Sarah Hörst for helpful discussions, as well as an anonymous referee for a thoughtful review of our work.

**Figures.**

Figure 1. OPUS preview images illustrating good coverage over Iapetus' dark material for three observations used in this study: the yellow cross denotes the position of the sun, and the white cross indicates the overhead location of the spacecraft.
(a) COCIRS_5709/DATA/APODSPEC/SPEC0709101715_FP3,
(b) COCIRS_5412/DATA/APODSPEC/SPEC0412310235_FP3,
(c) COCIRS_5702/DATA/APODSPEC/SPEC0702131700_FP3

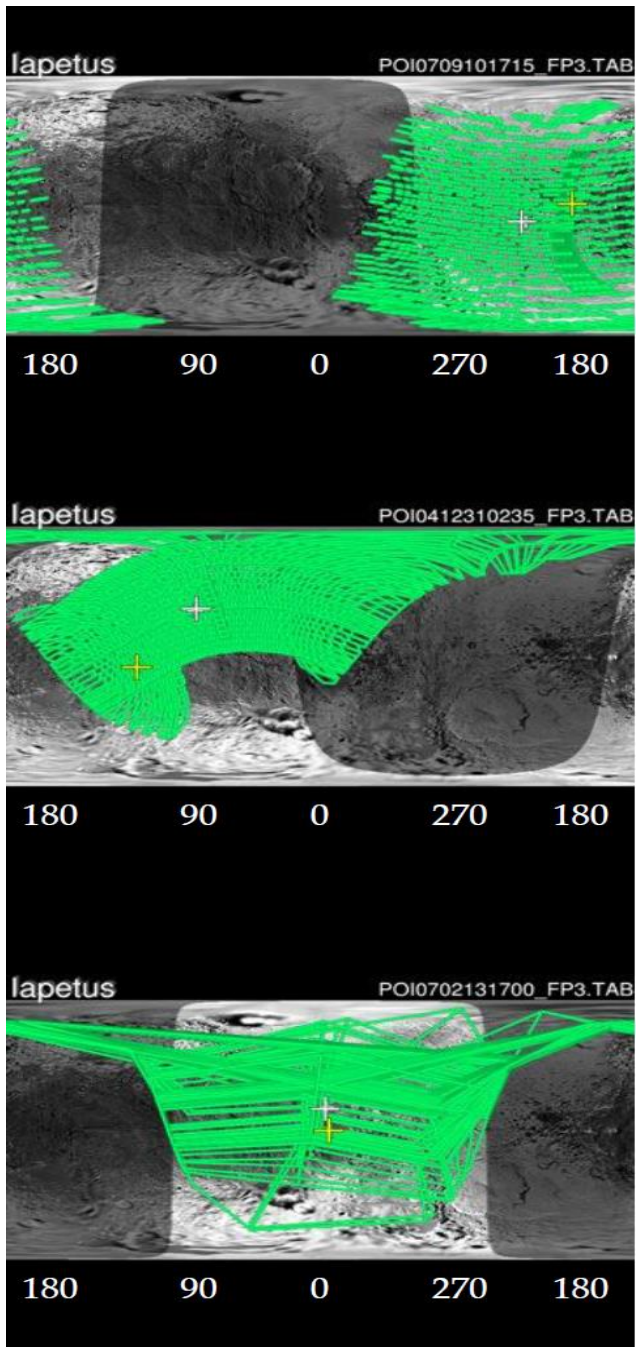

Figure 2. Boxed domains for three observations on Iapetus. There were 46 spectra collected in the orange box (centered at ~165ºW), five spectra in the purple box (centered at ~45ºW), and 1 spectrum in the red box (centered at ~310ºW) from Figure one observations a, b, and c, respectively. The field of view average emission angle ranges for spectra averaged from each observation are as follows: 49º – 81º (orange box), 34º – 45º (purple box), and 28º (red box). The surfaces constrained by the orange box tended to include intermediate elevations, except for the high elevation area located along the ridgeline at 0º latitude, while the purple and red boxes tended to only contain areas of high elevation that were not located directly on the ridgeline (Giese et al. 2008). Larger areas similar to the orange box were not averaged in the purple and red box regions because this tended to cause any features to vanish. The reasons for this are enigmatic, but are perhaps due to the differences in surface compositions between topographic features, the proximity to bright material in each of the regions, and/or changes in spacecraft geometry between the three observations.

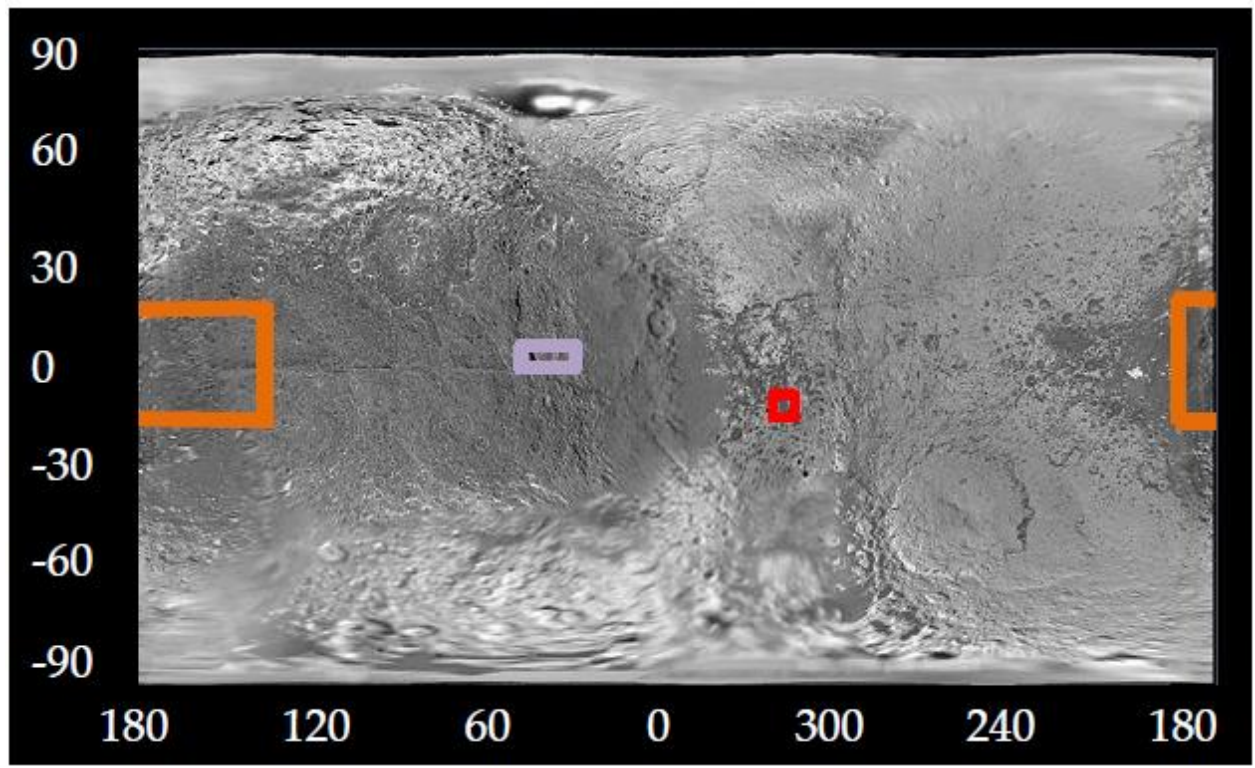

Figure 3. (a) Individual calibrated spectra for five detectors at one location (~7°S, 220°W). The detector number for each spectrum is indicated in the legend. (b) The average of the five spectra illustrating the improvements in S/N that result from averaging spectra compared to unaveraged spectra. Orange arrows denote regions of known instrument artifacts. (c) Radiance curves for colored boxed regions from Fig. 2 produced by averaging all spectra within a region, a blackbody at 126 K, and deep space for reference. Radiance features at 855 cm$^{-1}$ are denoted by an orange arrow and do not correspond to any known artifact.

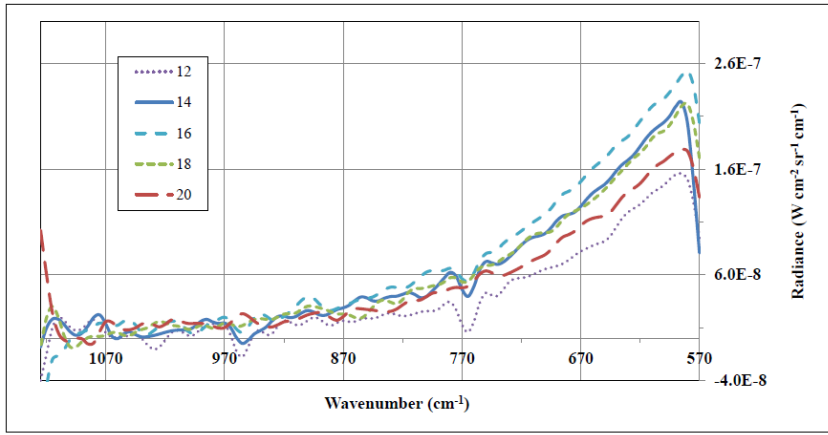

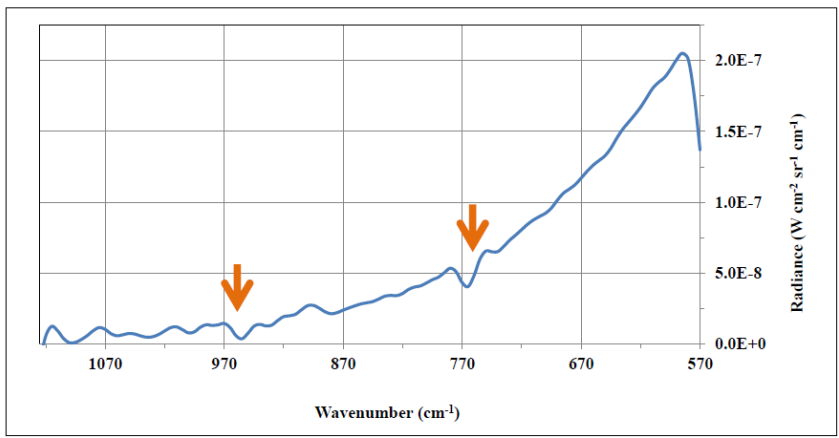

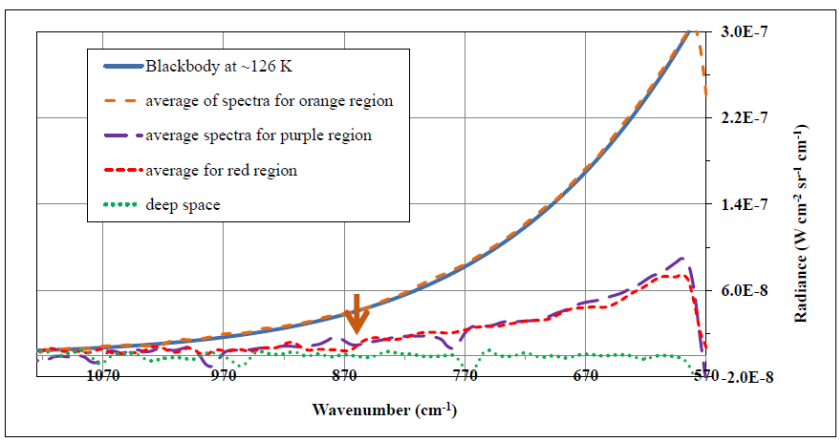

Figure 4. Emissivity curve generated from the average CIRS radiance spectrum of the orange boxed region and a blackbody at 126 K. A feature of interest is marked by the black line at 855 cm$^{-1}$ and is not an artifact. In addition, a possible doublet is detected at ~660 and ~690 cm$^{-1}$. The ~840 cm$^{-1}$ feature on Mars is attributed to fine-grained silicate minerals such as feldspars (Christensen et al. 2004). Very fine grains are required to produce this "transparency feature" along with a lack of silicate reststrahlen bands. Antigorite and lizardite spectra showing the best matches from published data to the CIRS feature at 855 cm$^{-1}$ (Bishop et al. 2008a, 2008b). Antigorite and lizardite also exhibit features at ~660 cm$^{-1}$ that may represent part of the detected doublet. Muscovite measured at 125 K in vacuum (2.7x10$^{-8}$ torr), vertically offset for clarity, also exhibits features at 855 cm$^{-1}$ and in the region of the doublet. It is noted that measured spectra have been converted to emissivity by one minus the spectral reflectance.

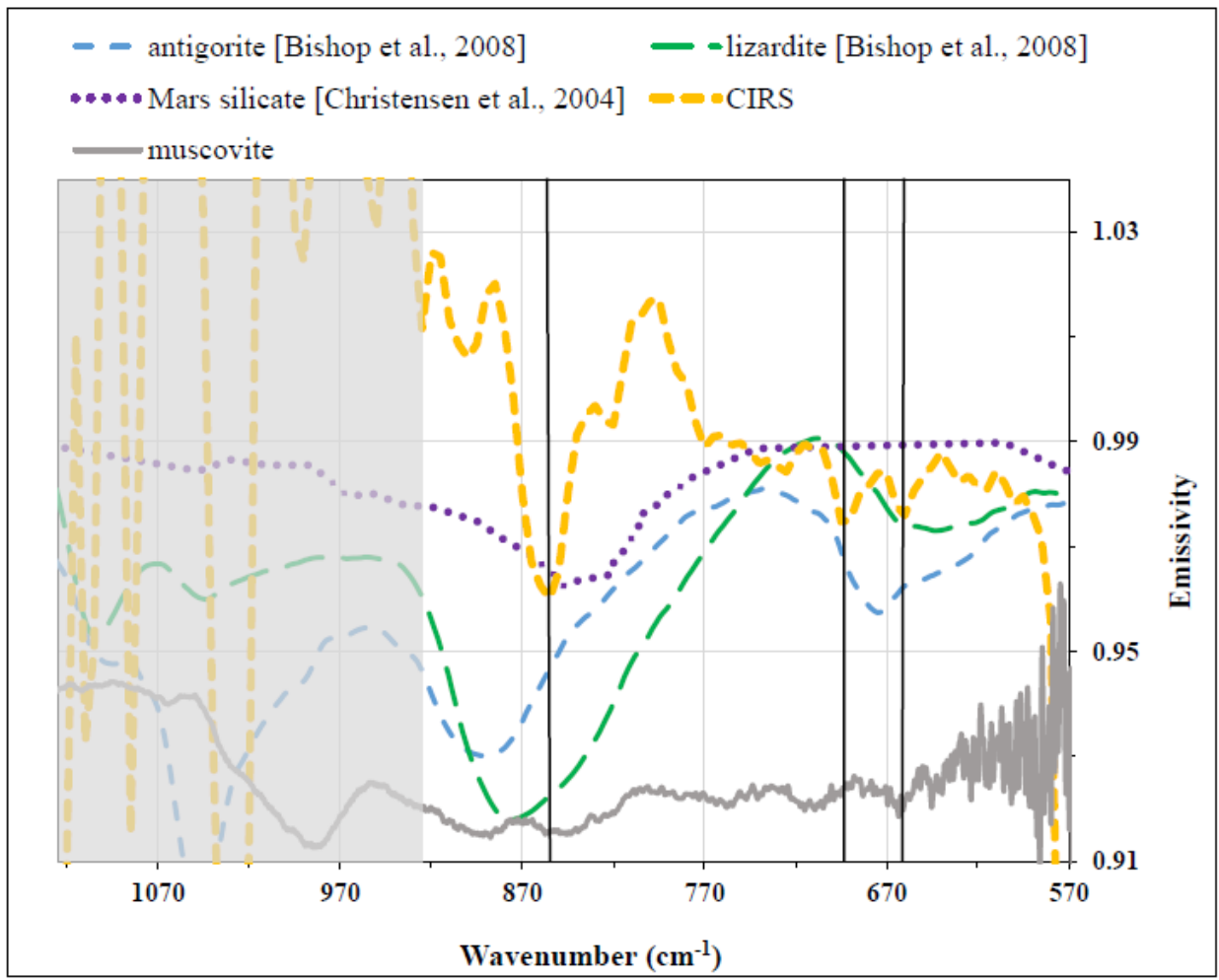

# References

Allodi, M. A., Ioppolo, S., Kelley, M. J., et al. 2014, Phys Chem Chem Phys, 16, 3442

Bishop, J. L., Dyar, M. D., Sklute, E. C., & Drief, A. 2008a, Clay Minerals, 43, 55

Bishop, J. L., Lane, M. D., Dyar, M. D., & Brown, A. J. 2008b, Clay Minerals, 43, 35

Buratti, B. J., Hicks, M. D., Tryka, K. A., Sittig, M. S., & Newburn, R. L. 2002, Icar, 155, 375

Buratti, B. J., Cruikshank, D. P., Brown, R. H., Clark, R. N., et al. 2005, ApJ, 622, L149

Carvano, J. M., A. Migliorini, A. Barucci, M. Segura, et al. 2007, Icar, 187, 574

Christensen, P. R., Wyatt, M. B., Glotch, T. D., Rogers, A. D., et al. 2004, Sci, 306, 1733

Clark, R. N. 1999, in Manual of Remote Sensing, Vol. 3, ed. A. N. Rencz, (New York, NY: John Wiley and Sons), 58

Clark, R. N., Brown, R. H., Jaumann, R., Cruikshank, D. P., et al. 2005, Natur, 435, 66

Clark, R. N., Swayze, G. A., Wise, R. A., Livo, K. E., et al. 2007, USGS digital spectral library splib06a, (Denver, CO:U.S. Geological Survey), Digital Data Series 231, http://speclab.cr.usgs.gov/spectral.lib06.

Clark, R. N., Curchin, J. M., Jaumann, R., Cruikshank, D. P., et al. 2008, Icar, 193, 372

Clark, R. N., Cruikshank, D. P., Jaumann, R., Brown, R. H., et al. 2012, Icar, 218, 831

Clark, R.N., Perlman, Z., N., Pearson, & Cruikshank, D. P. 2014, DPS Meeting 46 (Tucson, AZ), Abstract #502.06, http://adsabs.harvard.edu/abs/2014DPS....4650206C

Cruikshank, D. P., Bell, J. F., Gaffey, M. J., Hamilton Brown, R., et al. 1983, Icar, 53, 90

Cruikshank, D. P., Meyer, A. W., Brown, R. H., Clark, R. N., et al. 2010, Icar, 206, 561

Dalton, J. B., Prieto-Ballesteros., O., Kargel, J. S., Jamieson, C. S., et al. 2005, Icar, 177, 472

Dalton, J. B., III, Cruikshank, D. P, & Clark, R. N. 2012, Icar, 220, 752

Flanner, M. G., Gardner, A. S., Eckhardt, S., Stohl, A., et al. 2014, JGR: Atmos, 119, 9481

Flasar, F. M., Kunde, V. G., Abbas, M. M., Achterberg, R. K., et al. 2004, Space Sci. Rev., 115, 169

Giese, B., Denk, T., Neukum, G., Roatsch, T., et al. 2008, Icar, *193*, 359

Hand, K. P., C. F. Chyba, J. C. Priscu, R. W. Carlson, et al. 2009, in Europa, ed. R. T. Pappalardo, W. B. McKinnon, and K. Khurana (Tucson, AZ: Univ Arizona Press), 589–629

Howett, C. J. A., Spencer, J. R., Pearl, J., & Segura, M. 2010, Icar, 206, 573

Howett, C., Spencer, J., Hurford, T., Verbiscer, A., et al. 2014, AGU Fall Meeting (San Francisco, CA), Abstract #P43B-3987, http://adsabs.harvard.edu/abs/2014AGUFM.P43B3987H

Hsu, H. W., Postberg, F., Sekine, Y., Shibuya, T., et al. 2015, Natur, 519, 207

Jarvis, K. S., Vilas, F., Larson, S. M., & Gaffey, M. J. 2000, Icar, 146, 125

Johnson, A. P., Cleaves, H. J., Dworkin, J. P., Glavin, D. P., et al. 2008, Sci, 322, 404

Kempf, S., Srama, R., Postberg, F., Burton, M., et al. 2005, Sci, 307, 1274

Lebofsky, L. A., Feierberg, M. A., & Tokunaga, A. T. 1982, Icar, 49, 382

Logan, L. M., Hunt, G. R., Salisbury, J. W., & Balsamo, S. R. 1973, JGR, 78, 4983

Martin, W., Baross, J., Kelley, D., & Russell, M. J. 2008, Nature Reviews Microbiology, 6, 805

McAdam, M. M., Sunshine, J. M., Howard, K. T., & McCoy, T. M. 2015, Icar, 245, 320

Neugebauer, G., Matthews, K., Nicholson, P. D., Soifer, B. T., et al. 2005, Icar, 177, 63

O'Connor, B. H., & Chang, W. J. 1986, X-Ray Spectrometry, 15, 267

Palmer, E. E., & Brown, R. H. 2011, Icar, 212, 807

Pinilla-Alonso, N., Roush, T. L., Marzo, G. A., Cruikshank, D. P., et al. 2011, Icar, 215, 75

Presley, M. A., & Christensen, P. R. 1997, JGR, 102, 6551

Rivera-Valentin, E. G., Blackburn, D. G., & Ulrich, R. 2011, Icar, 216, 347

Salisbury, J. W., & Walter, L. S. 1989, JGR, 94, 9192

Salisbury, J. W., D'Aria, D. M., & Jarosewich, E. 1991, Icar, 92, 280

Sekine, Y., Shibuya, T., Postberg, F., Hsu, H.-W., et al. 2013, DPS Meeting 45 (Denver, CO), Abstract #403.02, http://adsabs.harvard.edu/abs/2013DPS....4540302S

Shirley, J. H., Kamp, L. W., & Dalton, J. B. 2013, AGU Fall Meeting (San Francisco, CA), Abstract #P54A-07, http://adsabs.harvard.edu/abs/2013AGUFM.P54A..07S

Soter, S. 1974. Brightness of Iapetus. Paper presented at IAU Colloq. 28, Cornell University, August 1974.

Smith, B. A., Soderblom, L., Batson, R., Bridges,P., et al. 1982, Sci, 215, 504

Spencer, J. R. 1987, The surfaces of Europa, Ganymede, and Callisto: An investigation using Voyager IRIS thermal infrared spectra, Ph.D. Thesis, University of Arizona, Tucson, AZ, 211

Stephan, K., Jaumann, R., Wagner, R., Clark, R. N., et al. 2010, Icar, 206, 631

Stephan, K., Jaumann, R., Wagner, R., Clark, R. N., et al. 2012, Space Sci., 61, 142

Tosi, F., Turrini, D., Coradini, A., Filacchione, G., et al. 2010, Mon. Not. R. Astron. Soc., 403, 1113

White, D. W., Gerakines, P. A., Cook, A. M., & Whittet, D. C. B. 2009, ApJS, 180, 182

Wray, J. J., Young, C. L., Hand, K. P., Poston, M. J., et al. 2014, In DPS Meeting 46 (Tucson, AZ), Abstract #418.16, http://adsabs.harvard.edu/abs/2014DPS....4641816W

Young, C. L., Sokolik, I. N., Flanner, M. G., & Dufek, J. 2014, JGR: Atmos, 119, 11387